\documentclass[aps,11pt,preprintnumbers]{revtex4-1}
\usepackage{graphicx}
\usepackage[english]{babel}
\usepackage{amsmath,amssymb,bm}
\usepackage[mathscr]{eucal}
\usepackage[colorlinks,linkcolor=blue,citecolor=blue]{hyperref}
\usepackage{yfonts}

\newif\ifarxiv
\arxivfalse

\begin		{document}
\def\st    {\begin{equation}}
\def\stp    {\end{equation}}

\def\tauo		{ \tau_o}

\def\half	{{\textstyle \frac 12}}

\def\Arxiv      #1 [#2]{\href{http://arxiv.org/abs/#1}{{\tt arXiv:#1 [#2]}}\,}

\def\half{{\textstyle\frac{1}{2}}}

\definecolor{Blueberry}{rgb}{0.25,0,0.65}
\definecolor{Strawberry}{rgb}{0.65,0,0.25}
\newcommand{\HL}{}
\newcommand{\PC}{}

\newcommand{\w}{{\textswab{w}}}

\newcommand{\be}{\begin{equation}}
\newcommand{\ee}{\end{equation}}
\newcommand{\bea}{\begin{eqnarray}}
\newcommand{\eea}{\end{eqnarray}}
\newcommand{\ben}{\begin{enumerate}}
\newcommand{\een}{\end{enumerate}}

\newcommand\ep{\epsilon}

\newcommand\Lam{\Lambda}

\newcommand\vep{\varepsilon}
\newcommand\ov{\over}
\newcommand\ha{{\half}}
\def\le{\left}
\def\ri{\right}

\newcommand\fluc{\zeta}

\preprint{MIT-CTP 4565}

\title
    {
   Defect formation beyond Kibble-Zurek mechanism and holography
    }
\author {Paul~M.~Chesler}
\affiliation{Department of Physics, Harvard University, Cambridge MA 02139, USA}
\email	{pchesler@physics.harvard.edu}
\author{Antonio M. Garc\'{\i}a-Garc\'{\i}a}
\affiliation{TCM Group, Cavendish Laboratory, University of Cambridge, JJ Thomson Avenue, Cambridge, CB3 0HE, UK}
\email       {amg73@cam.ac.uk}
\author	{Hong~Liu}
\affiliation {Center for Theoretical Physics, MIT, Cambridge MA 02139, USA}
\email      {hong\_liu@mit.edu}


\date{\today}

\begin{abstract}
We study the dynamic after a smooth quench across a continuous transition from the disordered phase to the ordered phase. Based on scaling ideas, linear response and the spectrum of unstable modes, we develop a theoretical framework, valid for any second order phase transition, for the early-time evolution of the condensate in the broken phase. Our analysis unveils a novel period of non-adiabatic evolution after the system passes through the phase transition, where a parametrically large amount of coarsening occurs before a well-defined condensate forms. Our formalism predicts a rate of defect formation parametrically smaller than the Kibble-Zurek prediction and yields a criterion for the break-down of Kibble-Zurek scaling for sufficiently fast quenches. We numerically test our formalism for a thermal quench in a 2 + 1 dimensional holographic superfluid. These findings, of direct relevance in a broad range of fields including cold atom, condensed matter, statistical mechanism and cosmology, are an important step towards a more quantitative understanding of dynamical phase transitions.
\end{abstract}


\maketitle

\section{Introduction}

Driving a system smoothly from a disordered to 
an ordered phase unveils the rich, and still poorly understood, phenomenology 
of dynamical phase transitions, a research theme of interest in vastly different fields. 
The Kibble-Zurek (KZ) mechanism (KZM) describes the spontaneous generation of 
topological defects when a system is driven through a second order phase transition 
into the ordered phase \cite{kibble,Kibble:1980mv, zurek}. Numerical simulations
~\cite{
Laguna:1996pv,
PhysRevLett.101.115701,
num,
PhysRevA.76.043613,
PhysRevLett.104.160404,
das2012winding,
kolodrubetz2012nonequilibrium,
wang2014quantum,
xu2014quantum,
del2010structural,
mathey2010light} 
have confirmed the spontaneous generation of defects and the scaling exponent of the defect density with the quench rate predicted by the KZM.
The KZM has also been generalized to quantum phase
transitions~\cite{zurekqpt,spinchainexact,simplest} and has been employed to compute correlation functions \cite{gubser} in the scaling region.  

Different experiments, with ion-crystals~\cite{pyka2013topological,kzion}, ultra cold atomic gases~\cite{coldexp,coldexp2,lamporesi2013spontaneous}, 
spin-liquids~\cite{spinexp}, superconducting films~\cite{maniv2003observation}, polariton superfluids~\cite{lagoudakis2011probing}, Josephson junctions~\cite{carmi2000observation} and 
Helium~\cite{ruutu1996vortex,ruutu1998defect,helium3a}, have observed, with different levels of certainty, defect generation but a really quantitative comparison with the predictions of the KZM is still missing.

Let us briefly review the KZM \cite{zurek1996cosmological,dziarmaga2010dynamics,delCampo:2013nla}.
Consider a system with 
a second order phase transition at temperature $T_c$, below which 
a symmetry is spontaneously broken and  
an order parameter $\psi$ develops a condensate.  
In equilibrium at temperature $T > T_c$ the correlation length $\xi_{\rm eq}$
and relaxation time $\tau_{\rm eq}$ are related to the reduced temperature $\epsilon \equiv 1 - \frac{T}{T_c}$
by 
\begin{align}
\label{eq:equilibrium}
&\xi_{\rm eq} =   \xi_s  | \epsilon|^{-\nu}, &
\tau_{\rm eq} =   \tau_s | \epsilon|^{-\nu z}, &
\end{align}
for some scales $  \xi_s$, $  \tau_s$ and critical exponents $\nu$, $z$.  
Consider a  quench from $T_i > T_c$ 
to $T_f < T_c$ with quench protocol 
\begin{align}
\label{eq:linquench}
&\epsilon(t) = t/\tau_Q, & t \in (t_i, t_f),&
\end{align}
where $ t_i = (1 - T_i/T_c) \tau_Q < 0$ and $t_f = (1 - T_f/T_c) \tau_Q > 0$.
The system can respond adiabatically to the change in temperature until
$\tau_{\rm eq}(t) \sim |t|$.  This condition defines the freeze-out time
and length scale
\begin{align}
\label{eq:xiandtfreeze}
&t_{\rm freeze} =  \tau_s \textstyle \left (\frac{\tau_Q}{ \tau_s} \right )^{ \frac{\nu z}{1 + z \nu}}, &
\xi_{\rm freeze} =  \xi_s \textstyle \left (\frac{\tau_Q}{ \tau_s} \right)^{ \frac{\nu}{  1 + \nu z }} \ .&
\end{align}
During the interval $t \in (-t_{\rm freeze},t_{\rm freeze})$ the evolution of the system is 
essentially frozen.
\footnote
  { 
  Strictly speaking one should distinguish $t_{\rm freeze}^>$ for $T > T_c$ 
  and $t_{\rm freeze}^<$ for $T < T_c$ as they can differ by an $O(1)$ constant. We will suppress such differences for notational simplicities. 
  }
The correlation length $\xi_{\rm freeze}$ 
then imprints itself on the state at $t = + t_{\rm freeze}$.

The density of topological defects generated across the phase transition can then be estimated as 
\begin{equation}
\label{eq:KZdefectdensity}
\rho_{\rm KZ}  \sim 1/\xi^{d-D}_{\rm freeze} \sim \tau_Q^{ (d-D) \nu /(1 + \nu z )}, 
\end{equation}
where $d$ is the number of the spatial dimensions and $D$ is the number of dimensions of a defect. 
While the KZM is only supposed to determine the density of defects up to an O(1) factor, 
it often significantly overestimates the real density of defects observed in numerical calculations: 
one needs a ``fudge'' factor $f$ multiplying $\xi_{\rm freeze}$ with $f = O(10)$ \cite{delCampo:2013nla}. 
See also~\cite{PhysRevE.81.050101} for a recent discussion.  

One motivation of this paper is to develop a formalism for describing the growth and coarsening of 
the order parameter after $t_{\rm freeze}$ in a general critical system.   
Our analysis stresses a period of non-adiabatic evolution, 
before a well-defined condensate forms, where the system 
coarsens and the correlation length grows parametrically larger than $\xi_{\rm freeze}$. In particular, we show that in many systems, including conventional superconductors and superfluid $^4{\rm He}$, there could be a large logarithmic hierarchy between $t_{\rm freeze}$
and the time scale we refer to $t_{\rm eq}$ when one can sensibly measure the density of defects.
Thus our analysis reconciles the need
for a ``fudge" factor $f$. Moreover, our analysis yields a new criterion for the breaking of the KZ scaling (\ref{eq:KZdefectdensity}).
Our discussion can also be applied without essential changes
to quantum phase transitions. For definiteness, we will restrict discussion to thermal phase transitions 
throughout the paper.  

A second motivation of this paper is  to test  the scaling  
predicted by the KZM and its refinement  in strongly coupled systems using holographic duality. 
Holography  equates certain systems of quantum matter without gravity to classical 
gravitational systems in one higher spatial dimension ~\cite{maldacena,witten,Gubser:1998bc}. 
Hence complicated many-body physics can be mapped onto 
a solvable numerical gravity problem.  
Some examples include~\cite{Chesler:2013lia,Casalderrey-Solana:2013sxa,Adams:2012pj,Bantilan:2012vu,Chesler:2010bi,Beuf:2009zz,Chesler:2008hg,murata2010,sonner,agg2013} (see also~\cite{Das:2011nk,Basu:2013soa} for a discussion of KZM for a holographic quantum quench).  
In this paper we study the KZM in a holographic superfluid in $2+1$ spacetime dimensions.
Our gravity calculation will provide a first check of KZ scaling 
in a strongly coupled system without quasiparticles and will verify key features of the 
coarsening physics discussed in the next section.

{\bf Note:} Independently, Sonner, del Campo and Zurek 
\cite{Sonner:2014tca} have found universal scaling behavior in the 
dynamics of strongly-coupled superconductors with a holographic dual.

\section{Far-from-equilibrium  
    coarsening }
    
\subsection{Unstable critical modes}    
    
 We now develop a formalism to 
describe a period of non-adiabatic growth of the order parameter $\psi$ after $t_{\rm freeze}$.
The seeds for condensate growth come from thermal and quantum fluctuations,
whose effects on the macroscopic evolution 
of $\psi$ can be described in terms of a stochastic source $\varphi$ for $\psi$. 
In the IR the statistics of the fluctuations in $\varphi$ read 
\begin{equation}
\label{nne}
\langle  \varphi^* (t, \bm x) \varphi (t', \bm x')  \rangle = \zeta \delta (t-t') \delta (\bm x - \bm x'), 
\end{equation}
where $\fluc$ is a (weakly) temperature dependent constant.  


Let $\psi(t,\bm q)$ and $\varphi(t,\bm q)$
be the Fourier transformed order parameter and noise respectively.
At early times $\psi(t,\bm q)$ is small and can be described 
by linear response, 
\begin{equation}
\psi(t,\bm q) = \int dt' G_{\rm R}(t,t',q) \varphi(t',\bm q),
\end{equation}
where $G_R(t,t',q)$ is the retarded $\psi$ correlator.  Statistical homogeneity and isotropy imply 
$G_R$ only depends on $q = |\bm q|$.
The regime of validity for the linear response will be discussed below.
To elucidate the growth of $\psi$ and to extract the time evolution of the correlation length after $t_{\rm freeze}$, we study the evolution of the correlation function
\begin{equation}
\label{eq:Cdef}
C(t, \bm r) \equiv
\langle \psi^*(t, \bm x + \bm r)   \psi(t,\bm x) \rangle.
\end{equation}
Averaging over the noise (\ref{nne}) we find 
\begin{equation}
\label{eq:ClinResponse}
C(t, q) =  \int dt'  \, \fluc |G_{R}(t,t',q)|^2.
\end{equation}

As the \HL{dynamics} is essentially frozen between $(-t_{\rm freeze}, t_{\rm freeze})$, 
at  $t \sim t_{\rm freeze}$, the system is in a supercooled state for which the leading \PC{time dependence} of  $G_R$ can be obtained by analytically continuing  to below $T_c$ the equilibrium retarded correlator $G_{\rm eq}$ above $T_c$.  Close to $T_c$ the \PC{time dependence of}  $G_{\rm eq} (t, q)$
should be dominated by the leading pole $\w_0 (q)$ (the critical mode) of $G_{\rm eq} (\omega, q)$ in the complex frequency plane, i.e. 
\begin{align}
\label{bne}
&G_{\rm eq}(t, q) =  \theta(t)  H(q) e^{- i \w_0(\epsilon,q) t},& 
\w_0(\epsilon, q) =  \epsilon^{z \nu} h( q \epsilon^{-\nu}),&
\end{align}
where $H (q)$ is some function which depends weakly on $q$.
$h (x)$ is a universal scaling function which is analytic in $x^2$ for small $x$.
For $T>T_c$, $\w_0 (q,T)$ 
lies in the lower half $\omega$-plane, and its imaginary part at $q=0$ gives the inverse 
of the relaxation time.%
\footnote
  {
  In the language of the dual gravitational description discussed below, $\w_0$ is the lowest quasinormal mode frequency of a dual black hole.
  }
When continued to a supercooled state at $T < T_c$, $\w_0$ moves to the 
upper half frequency plane for $q$ smaller than a certain $q_{\rm max}$, and for such $q$'s ~\eqref{bne} grows 
exponentially with time. More explicitly, for positive $\epsilon$ we can expand ${\rm Im} \, \w_0 $ in small $q$ as 
\begin{equation}
\label{eq:quasinormaldisp}
{\rm Im} \, \w_0 = -a \epsilon^{(z - 2) \nu} q^2 + b \epsilon^{z \nu} + \dots,
\end{equation}
where $a$ and $b$ are positive constants.  Hence 
${\rm Im}\, \w_0 > 0$ until $q \sim q_{\rm max}$ with
\begin{equation}
q_{\rm max} \sim \epsilon^{\nu}.
\end{equation}
Now let us consider the limit of slow quenches $\epsilon'(t) \to 0$. Assuming that the Green function depends weakly on temperature,
then for a short interval $t - t' \ll {1 \ov |\w_0|}$ \eqref{bne} should still apply, if $\w_0$ changes with 
time sufficiently slow, i.e.  
\be 
|\partial_t \log \w_0 (T(t))| \ll |\w_0| .
\ee

Under this approximation, $G_R$ then satisfies a first order differential equation 
\be \label{on}
\partial_t G_R (t,t';{\vec k}) = - i \w_0 (T (t)) G_R (t,t'; ) + {\vec k}\cdots , \qquad t > t' \
\ee
the integration of which leads to 
\begin{equation}
\label{eq:gradcorr}
G_{R}(t,t',q) = \theta(t-t')  H(q) e^{- i \int^t_{t'} dt'' \w_0(\epsilon(t''),q)  } \ .
\end{equation}
From $|\partial_t \log \w_0| < |\w_0|$, and using \eqref{eq:quasinormaldisp} and \eqref{eq:xiandtfreeze}, it is straightforward to show the earliest time when (\ref{eq:gradcorr}) can be applied is precisely $t > t_{\rm freeze}$. 

Substituting (\ref{eq:gradcorr}) into (\ref{eq:ClinResponse}) we then  
secure 
\begin{align}
\label{eq:Cint}
C(t,q) =  \int_{t_{\rm freeze}}^t  dt' \fluc |H(q)|^2  e^{2  \int^t_{t'} dt'' {\rm Im} \, \w_0(\epsilon(t''),q)} 
+ \cdots \ .
\end{align}
\PC{The $\cdots$ in (\ref{eq:Cint}) denotes the contributions in (\ref{eq:ClinResponse}) coming from the integration domain $t' < t_{\rm freeze}$, which 
will be neglected in our discussion below as the first term in~\eqref{eq:Cint} grows exponentially with time 
and will soon dominate. 
\footnote
  {
  \PC{In addition to the exponential suppression in time, when Fourier transformed to real space the 
  omitted terms in (\ref{eq:Cint}) also fall off parametrically faster with distance than the first term.}
  }
We note that $\w_0$ can also have a real part and therefore lead to oscillations of the order parameter superimposed to the exponential growth induced by the imaginary part. This is an interesting phenomenon that deserves further discussion.
Let us consider the behavior of the above integral for 
$t$ parametrically large compared to $t_{\rm freeze}$ assuming for the moment that the linear analysis holds. 
For this purpose it is convenient to introduce 
\begin{equation}
\bar t \equiv  \frac{t}{t_{\rm freeze}}.
\end{equation}
In the regime $\bar t \gg 1$, we find for $q \xi_{\rm freeze} \ll 1$~(see Appendix~\ref{app:b} for details), 
\be \label{eq:Cfourierspace}
C(t,q) = a_1 \fluc t_{\rm freeze} \exp \left(a_2 \bar  t^{1 + \nu z} - {1 \over 2} q^2 \ell_{\rm co}^2 (\bar t) \right)  
\ee
where 
\be \label{eq:coarsening}
\ell_{\rm co} (\bar t)  = a_3 \xi_{\rm freeze} \bar t^{{1 + (z-2) \nu \over 2}} \ 
\ee
and $a_1, a_2, a_3$ are $O(\tau_Q^0)$ constants.
Fourier transforming $q$ to coordinate space we find 
\begin{equation}
\label{eq:Crealspace}
C(t,r) \sim |\psi|^2 (t) e^{-  {r^2 \over 2 \ell_{\rm co}^2 (t)} }, 
\quad {\rm with} \quad |\psi|^2 (t) \sim \tilde \varepsilon ( t)  
e^{a_2 \bar t^{\,1 + z \nu}}
\end{equation}
where 
\begin{equation} \label{eq:mm}
\tilde \varepsilon ( t) \equiv \fluc t_{\rm freeze} \ell_{\rm co}^{-d} (t) \sim  \varepsilon  t_{\rm freeze}  
\bar t^{ \, -{d(1+ \nu (z-2)) \over 2}} 
\end{equation}
with
\begin{equation}
\label{eq:mm1}
 \varepsilon \equiv  \fluc  \xi_{\rm freeze}^{-d} \sim \fluc \tau_Q^{-{d \nu \over 1 + \nu z}} \ .
\end{equation}
Equations~\eqref{eq:Cfourierspace}--\eqref{eq:mm1} are our main results of this section.  
We now proceed to discuss their physical meaning and physical implications.

\subsection{Equilibration time and density of defects}

With the usual inverse volume
dependence, $\varepsilon$ defined in~\eqref{eq:mm1} may be interpreted as the  effective parameter characterizing fluctuations for a spatial region of size $\xi_{\rm freeze}$,  while $\varepsilon t_{\rm freeze}$ \PC{may be interpreted} as the fluctuations accumulated 
over a time scale of order $O(t_{\rm freeze})$. In the limit of large $\tau_Q$, $\varepsilon$ goes to zero, justifying the use of linear response. 
The linear response analysis should break down at some point which can be estimated by 
comparing $|\psi|^2 (t)$ in~\eqref{eq:Crealspace} with the equilibrium value of the condensate square. 
\HL{Recall that in equilibrium, the expectation value of an order parameter for 
reduced temperature $\epsilon \ll 1$ is characterized by a critical exponent $\beta$} 
\be \label{eqv}
 |\psi|^2_{\rm eq} (\ep) \sim \epsilon^{2 \beta}  \ .
 \ee
Introducing a scale $t_{\rm eq}$ by requiring 
\be \label{teq}
|\psi|^2 (t = t_{\rm eq}) \sim  |\psi|^2_{\rm eq} (\ep (t_{\rm eq})),
\ee
we expect the linear analysis to break down for $t \sim t_{\rm eq}$. 
\PC{In particular, for $t \gtrsim t_{\rm eq}$, we expect the condensate growth to transition from the exponential
growth of~\eqref{eq:Crealspace} to the adiabatic growth governed by~\eqref{eqv} with 
$\epsilon$ in (\ref{eqv}) given by the time-dependent reduced temperature (\ref{eq:linquench}).}
\PC{Moreover, the system does not contain 
a well-defined number of topological defects until a well defined condensate forms which necessarily lies outside 
the domain of linear response}. Thus 
$t_{\rm eq}$ is also the natural time scale to measure the density of 
topological defects.  

To estimate $t_{\rm eq}$ we have to solve Eq. (\ref{teq}) for $t_{\rm eq}/t_{\rm freeze}$. From \eqref{eq:coarsening},\eqref{eq:Crealspace},  \eqref{eq:mm}, \eqref{eq:mm1}, \eqref{eqv},  
it is clear that this ratio, that determine the duration of the coarsening region, is controlled by the dimensionless parameter,
\be \label{defR}
\textstyle R \equiv \frac{\epsilon^{2 \beta}}{ \tilde \varepsilon ( t)} \simeq  {\tau_Q^{- {2 \beta \over 1 + \nu z}}  \over \vep  t_{\rm freeze}} \sim \fluc^{-1} \tau_Q^{{\Lam \ov 1+ \nu z}}, \qquad \Lam \equiv {(d - z)\nu -2 \beta} 
\ee
When $R \lesssim O(1)$ we must have $t_{\rm eq} \sim t_{\rm freeze}$. In this case there is no hierarchy of scales between $t_{\rm freeze}$ and $t_{\rm eq}$ and the condensate begins to grow adiabatically after $t_{\rm freeze}$.
In other words, in this case our analysis reduces to the standard story of the KZM and the density of topological defects is given by~(\ref{eq:KZdefectdensity}).
When $R \gg 1$ there is, however, a hierarchy between $t_{\rm eq}$ and $t_{\rm freeze}$, 
and~\eqref{eq:Crealspace} applies over a parametrically large interval of time during which 
the condensate grows with time exponentially, and  the coarsening
length $\ell_{\rm co} (t)$, which controls the typical size of a condensate droplet, grows with time as a power. 
\PC{In particular in the limit $R \to \infty$, from~\eqref{eq:Crealspace}, \eqref{teq} and \eqref{eq:coarsening} 
we see
\be \label{uuen}
\textstyle {t_{\rm eq} \ov t_{\rm freeze}} \sim (\log R)^{1 \ov 1 + \nu z} + \cdots \to \infty \ ,
\ee
and
\begin{equation}
\textstyle {\ell_{\rm co}(t_{\rm eq}) \ov \xi_{\rm freeze}} \sim (\log R)^{\frac{1 + (z - 2)\nu}{2(1 + z \nu)}} + \cdots \to \infty.
\end{equation}
Thus for $R \gg 1$  a parametrically large amount of coarsening occurs \textit{before} a well-defined condensate even forms.
The density of topological defects of dimension $D$ is then (using~\eqref{eq:coarsening})
\begin{equation}
\label{eq:scaling}
\rho (t_{\rm eq}) \sim 1/\ell_{\rm co}^{d - D} ( t_{\rm eq}) \sim 
 \left [\textstyle \log (\fluc^{-1} \tau_Q^{\Lam \ov 1+ \nu z}) \right ]^{-\frac{(d - D)(1 + (z - 2)\nu)}{2(1 + z \nu)}} \rho_{\rm KZ} .
\end{equation}
As a result of early-time coarsening, the defect density $\rho$ is parametrically much smaller than  Kibble-Zurek prediction $\rho_{\rm KZ}$ and 
the standard KZ scaling is corrected by a logarithmic prefactor. Possible systems with $R \gg 1$ will be further discussed in the conclusion section.}

We stress that the time dependence of~\eqref{eq:Crealspace} differs from the scaling behavior of standard coarsening physics~\cite{Bray:1994zz}, which applies only {\it after} the magnitude $|\psi|$ has achieved its equilibrium value.  The possible importance of early-time coarsening physics in the KZM has recently also been 
discussed in~\cite{PhysRevE.81.050101}, but it assumed the scaling behavior of standard coarsening physics
and thus is not compatible with our result.    
\subsection{Rapid quenches} \label{sec:iic}
By decreasing $\tau_Q$ (while keeping $T_i, T_f$ fixed), eventually  the  scaling (\ref{eq:scaling}) for the defect density must break down. 
In standard KZ discussions, this should happen when $t_f \lesssim t_{\rm freeze}$.
Here we point out that for systems with $t_{\rm eq} \gg t_{\rm freeze}$, the  scaling (\ref{eq:scaling}) breaks down
for $t_f \ll t_{\rm eq}$, and can happen even for $t_f$ parametrically much {\it larger} than $ t_{\rm freeze}$.
This is easy to understand; for $t_{\rm eq} \gg t_f \gg t_{\rm freeze}$, since the system stays at $T_f$ 
after $t_f$, the growth of the condensate will largely be controlled by the unstable modes 
at $T_f$, and the defect density will be determined by $T_f$ rather than $\tau_Q$. 
 We now generalize the above discussion of far-from-equilibrium coarsening to such a case, where 
equation~\eqref{eq:Cint} should be modified to 
\be
\label{eq:Cint2}
C(t,q) =  \int_{t_{\rm freeze}}^t  dt' \fluc |H(q)|^2  e^{2 {\rm Im} \, \w_0(\epsilon_f ,q) (t - t')}  + \cdots
\ 
\ee
\HL{where as commented below~\eqref{eq:Cint}, $\cdots$ denotes contributions from earlier times which 
can be neglected in subsequent discussions.}
Note $\w_0(\epsilon_f ,q)$ is now evaluated at $\ep_f \equiv {T_c - T_f \ov T_c}$ which results in 
a simple  $e^t$ growth for any $\nu, z$ (compare with~\eqref{eq:Cfourierspace}). 
Fourier transforming the above expression, then $C(t, r)$ can be written in a scaling form~(see Appendix~\ref{app:b} for details)
\be \label{nmm}
C(t,r) \sim \ep_f^{(d-z) \nu} \fluc  f(\ep_f^{\nu z} (t- t_{\rm freeze}), r  \ep_f^\nu)
\ee
for some scaling function $f$. For $\ep_f^{\nu z} (t-t_{\rm freeze}) \gg 1$ and $r  \ep_f^\nu \gg 1$  (assuming linear response still applies), $f$ can be obtained explicitly and one finds 
 \be \label{apme}
C (t, r) = |\psi|^2  (t) e^{- {r^2 \ov 2 \ell_{\rm co}^2 (t)}} , \qquad  |\psi|^2 (t) \sim \ep_f^{(d-z) \nu} \fluc  \exp \le[2 b  (t - t_{\rm freeze}) \ep_f^{\nu z}  \ri]
\ee
with 
\be 
\ell_{\rm co}^2 (t) = 4a  (t - t_{\rm freeze}) \ep_f^{\nu (z-2)} \ . 
\ee
Note that in comparing with~\eqref{eq:Cfourierspace}--\eqref{eq:coarsening}, we see that both 
the logarithm of the condensate square and the coarsening length square grow linearly with time. 
 
Parallel to the earlier discussion, we postulate that the linear response analysis breaks down when 
the condensate squared obtained from~\eqref{apme} becomes comparable to $|\psi|^2_{\rm eq}$. 
To estimate the time scale $t_{\rm eq}$ when this happens, it is again convenient to  
introduce 
\be \label{bmme}
\textstyle R_f \equiv {\ep_f^{2 \beta} \ov \fluc \ep_f^{(d-z) \nu} } = \fluc^{-1}  \ep_f^{- \Lam},
\ee
and the criterion for linear response to apply for $\ep_f^{\nu z} (t-t_{\rm freeze}) \gg 1$ is again 
$R_f \gg 1$. In particular, the equilibrium time $t_{\rm eq}$ and  
the density of defects should be given by 
\be \label{den}
t_{\rm eq} - t_{\rm freeze} \sim \begin{cases}
   \ep_f^{- z\nu}  & R_f \lesssim O(1) \cr
   \ep_f^{-  \nu z } \log R_f & R_f \gg 1 
   \end{cases} ,
\qquad
 \rho \sim \begin{cases}
   \ep_f^{(d-D) \nu}  & R_f \lesssim O(1) \cr
   \ep_f^{(d -D)\nu} \log^{-\frac{d-D}{ 2}} R_f & R_f \gg 1 
   \end{cases} \ .
\ee
Clearly $\rho$ is independent of $\tau_Q$. 

For very fast quenches, i.e. $t_f \ll t_{\rm freeze}$, the whole quench from $T_i$ to $T_f$ will 
be non-adiabatic. In such a case,  at the end of quench, the system will have correlation length $\xi_i \sim \xi_{\rm eq} (T_i)$  imprinted from the state before the starting of quench, and the scale  $t_{\rm freeze}$ is no longer relevant. But the above discussion of far-from-equilibrium coarsening still applies with $t - t_{\rm freeze}$ replaced by $t - t_f$.\footnote{Strictly speaking, the above discussion applies to $\xi_i < \xi_{\rm min} (T_f) \equiv q_{\rm max}^{-1} (T_f) \sim \ep_f^{-\nu}$. For $\xi_i >  \xi_{\rm min} (T_f) $,  unstable $q$ modes with $q^{-1} < \xi_i$ will be averaged out and only those modes with $q^{-1} > \xi_i$ can grow.}

\section{Numerical Results: out of equilibrium dynamic of a holographic superfluid}
\label{sec:gravdes}

In this section we test the predictions of the previous section by constructing numerical solutions for the time evolution of a $2+1$ dimensional holographic superfluid after a quench across a second order phase transition. 
In Appendix~\ref{app:a} we give a detailed account of the gravity setup and technical details. Here we discuss the main results.

\subsection{Predictions for holographic systems}

Many examples of field theories with gravity duals are now known in various spacetime
dimensions, which essentially consist of elementary bosons
and fermions interacting with non-Abelian gauge fields. The rank $N$ of the gauge group is
mapped to the Newton constant $G_N$ of the bulk gravity such that $G_N \sim {1 \ov N^2}$;
the classical gravity approximation in the bulk thus corresponds to the large $N$ limit in the boundary theory. 
Finite temperature in the boundary system is described on the gravity side by a black hole. 
In the large $N$ limit, thermal and quantum fluctuations are suppressed by $1/N^2$ and 
on the gravity side are encoded in quantum gravity effects induced from the black hole's Hawking radiation. 

In this paper we consider a holographic superfluid phase transition in two spatial dimension with relevant topological defects \PC{being} point-like vortices. In the large $N$ limit, the phase transition 
has mean field critical exponents with $\nu = \ha, z=2, \beta = \ha$,
and $\fluc$ in~\eqref{nne} of order $O( {1 \ov N^2})$.  
For such a system, the predictions from the KZM for density of superfluid vortices read 
\be 
t_{\rm freeze} \sim \tau_Q^{1 \ov 2} , \qquad \xi_{\rm freeze} \sim \tau_Q^{1 \ov 4}, \qquad \rho_{\rm KZ} \sim \tau_Q^{-{1 \ov 2}} \ . 
\ee
Applying the discussion of last section to such large $N$ theories, we 
can make the following predictions: 

\ben 

\item For slow quenches, i.e. quenches with $t_f \gg t_{\rm eq}$, with $d=2$ and mean field exponents, equation~\eqref{eq:Crealspace} becomes~($\bar t = t/t_{\rm freeze}$)
\be \label{menx}
C(t,r) \sim |\psi|^2 (t) e^{-  {r^2 \over 2 \ell_{\rm co}^2 (t)} }, 
\quad |\psi|^2 (t) \sim \tilde  \varepsilon  t_{\rm freeze} \bar t   e^{a_2 \bar t^2},
\quad \ell_{\rm co} ( t) \sim \xi_{\rm freeze}  \sqrt{\bar t } \ .
\ee
Furthermore, from~\eqref{defR} we find $\Lam = -1$ and thus 
\be 
R \sim \fluc^{-1} \tau_Q^{-\ha}  \sim N^2 \tau_Q^{-\ha} \ .
\ee
In the large $N$ limit, we always have $R \gg 1$ and from~\eqref{uuen} and \eqref{eq:scaling}
\be \label{prd1} 
{t_{\rm eq} \ov t_{\rm freeze}}  \sim {\textstyle \sqrt{\log {N^2 \ov \sqrt{\tau_Q}}}}\ ,
\ee
and
\be \label{prd2}
\rho  \sim {\textstyle \frac{1}{\sqrt{\log {N^2 \ov \sqrt{\tau_Q}}}} }  \rho_{\rm KZ} \ . 
\ee

\item For rapid quenches, with mean field exponents and $d=2$, equation~\eqref{apme} can be written as 
 \be \label{apme2}
C (t, r) = |\psi|^2  (t) e^{- {r^2 \ov 2 \ell_{\rm co}^2 (t)}} , \qquad  |\psi|^2 (t) \sim  \fluc  \exp \le[2 b  (t - t_{\rm freeze}) \ep_f  \ri]
\ee
with 
\be 
\ell_{\rm co}^2 (t) = 4a  (t - t_{\rm freeze})  \ . 
\ee
Furthermore $R_f = \fluc^{-1} \ep_f^{-1} \sim N^2 \ep_f^{-1}$ is always much greater than $1$ in the large $N$ limit, and we have 
from (\ref{den})
\be 
t_{\rm eq} - t_{\rm freeze} \sim 
\ep_f^{-1} \log {\textstyle  \frac{N^2}{\epsilon_f}}
\qquad
\rho \sim 
\frac{\ep_f}{\log {\textstyle  \frac{N^2}{\epsilon_f}}} 
\ . 
\label{rnw}
\ee
Note that both quantities above are independent of $\tau_Q$.

\een

\begin{figure}[h]
\includegraphics[scale = 0.48]{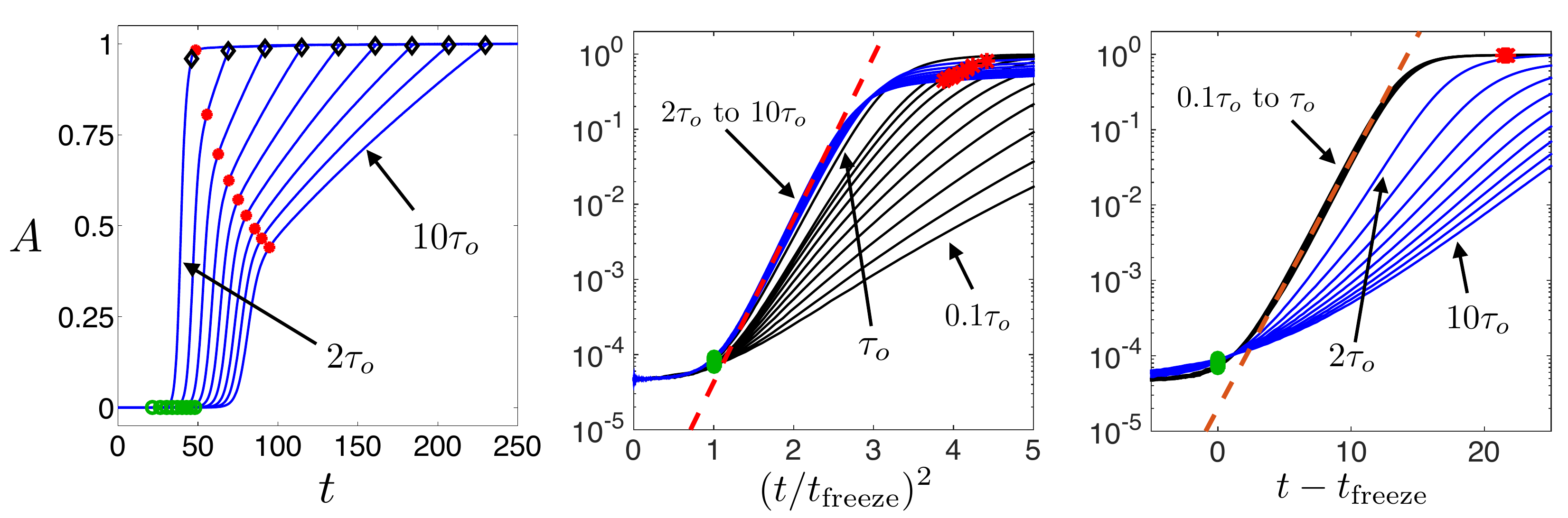}
\caption{{\it Left:} The normalized average condensate $A(t)$ defined in (\ref{eq:Adef}) for quench rates $\tau_Q = n \tauo$ for $n = 2,\dots,10$ (from left to right) with $\tauo = 161.37/T_c$.  
The black diamonds denote the times at which for each $\tau_Q$ the thermal quench is over. 
All curves  experience a period of rapid growth, which is followed by a period of approximate linear growth. 
The start of the rapid growth can be identified as $t_{\rm freeze}$, which 
we operationally define as the time at which $A(t) = 2 A(-\infty)$, and are denoted by 
the green circles. The crossover from exponential to linear growth corresponds to the equilibration time $t_{\rm eq}$ (\ref{teq}), which we operationally define as the time at which $A''(t_{\rm eq}) < 0.1\,  {\rm max} \left\{A''(t) \right\}$ and are labelled by the red stars.  {\it Middle:}  scaling behavior of slow quenches; when $A$ is plotted  v.s. $\bar t^2 = (t /t_{\rm freeze})^2$ only curves corresponding to slow quenches collapse into a single one. The linear behavior in the logarithmic 
plot is consistent with \eqref{menx}. {\it Right:} scaling behavior of rapid quenches; $A(t)$ is plotted v.s. $t - t_{\rm freeze}$ on a logarithmic scale  for quench rates $\tau_Q = n \tauo$ for $n = 0.1, 0.2 \dots,10$. As predicted from~\eqref{apme2} only $\tau_Q$ curves corresponding to fast quenches collapse into a single one, which exhibits linear exponential growth. All dimensional quantities are measured in units of $T_c$ (\ref{tcmu}).  
 \label{fig:AvgCond} } 
\end{figure}

\begin{figure}[h!]
\begin{center}
\vspace*{10pt}
\includegraphics[scale = 0.5]{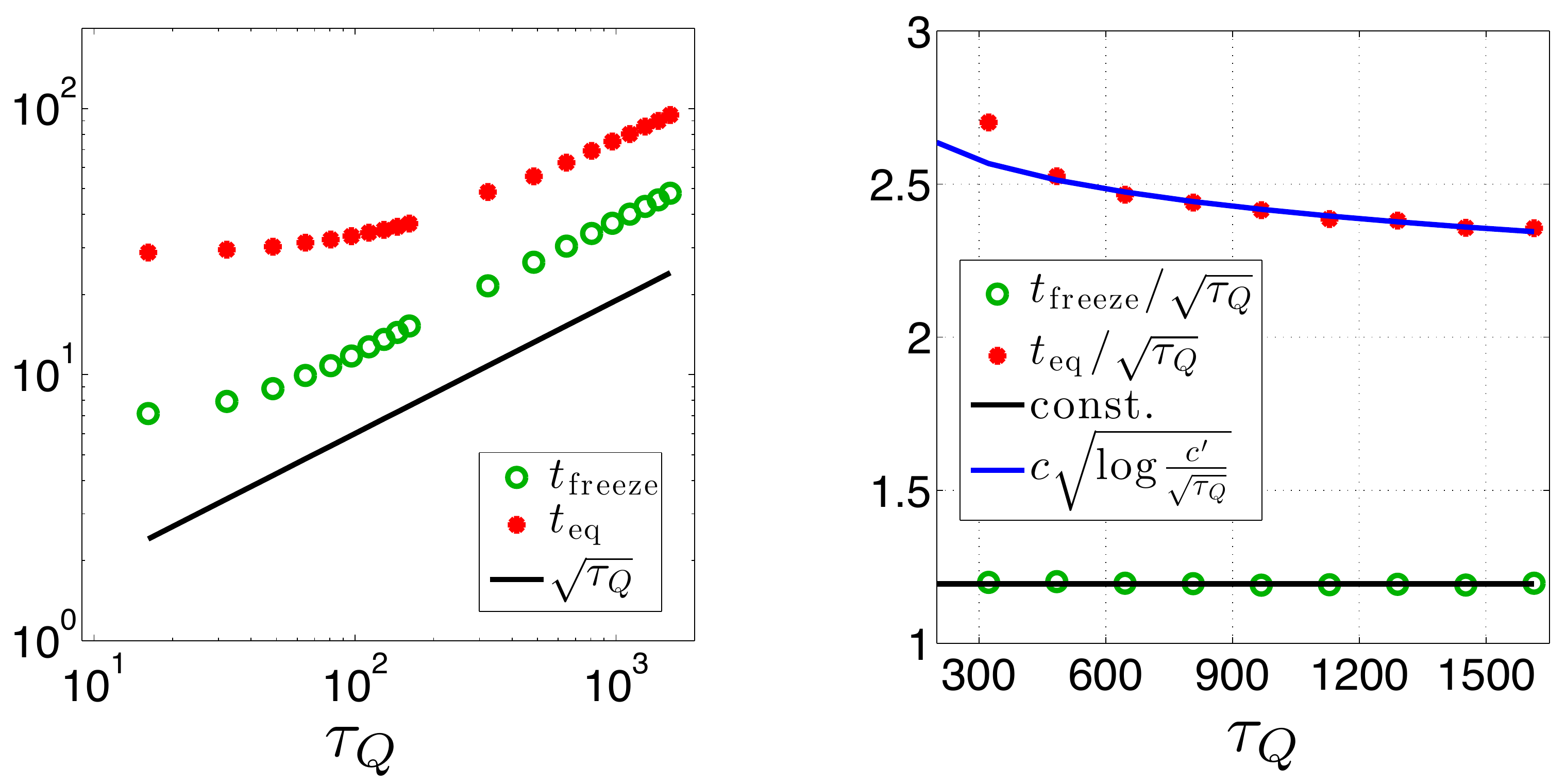}
\vspace*{-15pt}
\end{center}
\caption{ 
{\it Left:} The freeze-out time $t_{\rm freeze}$ and equilibration time  $t_{\rm eq}$ as a function of $\tau_Q$ expressed in units of $T_c$ (\ref{tcmu}).
For rapid quenches $t_{\rm eq} \to {\rm const}$, as expected.  For slow quenches both $t_{\rm freeze}$ and $t_{\rm eq}$ are consistent with the $\tau_Q^{1/2}$ scaling. 
{\it Right:}  $t_{\rm eq}$ also exhibits logarithmic correction to the $\tau_Q^{1/2}$ scaling, consistent with the prediction 
of~\eqref{prd1}.
\label{fig:teq} 
} 
\vspace*{-5pt}
\end{figure}

\subsection{Numerical results}

\HL{We have performed numerical simulations of thermal quenches across a second order phase transition 
of a $2+1$ dimensional holographic superfluid. 
We employ the linear quench~(\ref{eq:linquench}) which in the gravity context translates into a black hole with a time dependent temperature. Instead of directly computing fluctuations from Hawking radiation (see e.g.~\cite{CaronHuot:2011dr,Chesler:2012zk}), we model fluctuations from quantum gravity effects as a random noise that enters as a nontrivial boundary condition in the gravity equation of motion. In such a formulation, $\fluc$ can taken as an adjustable parameter which we take to be numerically small so as to imitate the $O(1/N^2)$ fluctuations. See
Appendix~\ref{app:a} for details.}  In what follows all dimensional quantities are expressed in units of the critical temperature $T_c$.

We begin our analysis by studying the normalized average order parameter 
\be \label{eq:Adef}
A(t) = \frac{1}{M}\sum_{i = 1}^M \frac{a_i(t)}{a_i(\infty)}, \qquad a_i(t) \equiv \int d^2 x \, | \psi_i(t,\bm x)  |^2  \ .
\ee
The sum over $i$ represents an ensemble average over $M$ configurations at fixed $\tau_Q$.   
The time evolution of $A(t)$ for various $\tau_Q$ are given in Fig.~\ref{fig:AvgCond}.
The left and middle plots correspond to slow quenches where we see 
all curves experience a period of rapid growth after $t_{\rm freeze}$
followed by a period of approximate linear growth. This regime of slow quenches is the one at which the KZ mechanism of defect formation applies. We note that the KZ mechanism assumes that defects are generated at $t_{\rm freeze}$ so it does not provide a theory of the condensate growth. However this problem has been previously addressed in the condensed matter literature \cite{stoof1999coherent,stoof2001dynamics}.
We operationally define $t_{\rm freeze}$ as the time at which $A(t) = 2 A(-\infty)$. 
 The rapid growth can be identified with the regime described by~\eqref{eq:Crealspace} and~\eqref{menx}
as indicated by the middle plot. 
  The linear growth can be identified as the regime of adiabatic condensate growth. 
To see this note that for mean field $|\psi|^2_{\rm eq} \sim \ep (t) = {t \ov \tau_Q}$ 
implying $A(t) \sim t/\tau_Q$ for adiabatic growth. This conclusion is supported by the 
slope of the linear growth, the observation that the termination of the linear condensate growth coincides with the end of the quench,
and that when extrapolated to $t = 0$ the linear curves have $A = 0$.
The crossover from exponential to linear growth corresponds to the equilibration time $t_{\rm eq}$ (\ref{teq}), which we operationally defined as the time in which $A''(t_{\rm eq}) < 0.1\,  {\rm max} \left\{A''(t) \right\}$.

A key feature of the middle plot of Fig.~\ref{fig:AvgCond} is that curves of different $\tau_Q$ all lie top of one another when we plot them 
in terms of scaling variable $\bar t^2 = (t/t_{\rm freeze})^2$.  In particular, the linear $\bar t^2$ growth in the logarithmic plot agrees very well with the prediction of~\eqref{menx}. A similar collapse for different $\tau_Q$'s was recently observed \cite{PhysRevLett.104.160404} in a $1D$ system governed by the stochastic Gross Pitaevskii equation.  
The right plot describes fast quenches discussed in Sec.~\ref{sec:iic}, with all the qualitative features of~\eqref{apme2} confirmed numerically, namely, $e^{t}$ growth as compared with the $e^{t^2}$ growth of slow quenches, and all curves of different $\tau_Q$ lying on top of one another when plotted v.s. $t - t_{\rm freeze}$. 
  For such a ``rapid'' quench, the growth of condensate and the resulting defect density are dictated by $T_f$ and are independent of $\tau_Q$. This expectation is also borne out in Fig.~\ref{fig:teq} where we plot the freeze-out time $t_{\rm freeze}$ and equilibration time  $t_{\rm eq}$ as a function of $\tau_Q$. While for large $\tau_Q$, their behavior is consistent with $\sqrt{\tau_Q}$ scaling, for rapid quenches, $t_{\rm eq}$ approaches a constant.  We note that in the context of condensed matter the formation of a one dimensional condensate was investigated numerically in Ref.\cite{stoof2001dynamics}.
The right panel of  Fig.~\ref{fig:teq} also shows that our numerical results are consistent with the presence of a logarithmic hierarchy between the two time scales as predicted in~\eqref{prd1}.

\begin{figure*}[ht!]
\includegraphics[scale = 0.55]{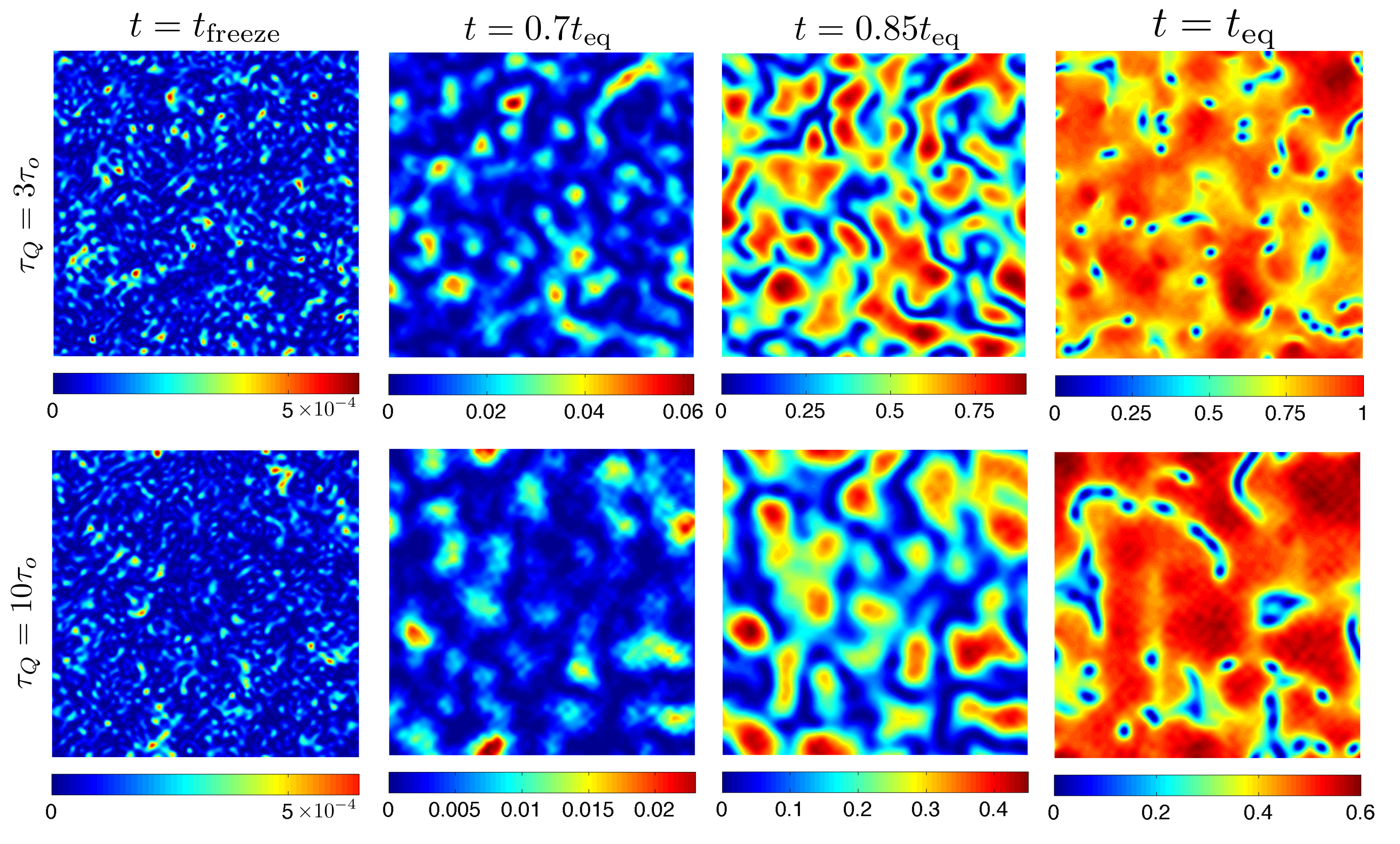}
\caption{The time evolution of 
$| \psi(t,\bm x) |^2/|  \psi(t = \infty,\bm x)  |^2$ for $\tau_Q = 3 \tauo$ (upper) and $\tau_Q = 10 \tauo$ (lower)
at times $t = t_{\rm freeze}$, $t = 0.7 \, t_{\rm eq}$, $t = 0.85 t_{\rm eq}$ and 
$t = t_{\rm eq}$. The key message is that we can sensibly talk about defect density only after $t_{\rm eq}$. 
At $t = t_{\rm freeze}$ the order parameter is very small and dominated by fluctuations. 
These fluctuations seed droplets of condensate, whose subsequent causal connection can be seen at time $t = 0.7 t_{\rm eq}$.
At such a time, the droplets are still separated by large regions where there is no condensate.  Subsequently, the droplets expand and grow in amplitude and the system becomes smoother and smoother.  By time $t_{\rm eq}$ the droplets 
have merged into a comparatively uniform condensate with isolated 
regions where $ \psi = 0$.  The non-uniformities --- the localized blue ``dots" --- 
are superfluid vortices with winding number $\pm 1$. All dimensional quantities are expressed in units of $T_c$ (\ref{tcmu}). 
\label{fig:CondEvo}
} 
\end{figure*}

In Fig.~\ref{fig:CondEvo} we plot the time evolution of 
$| \psi(t,\bm x) |^2/|  \psi(t = \infty,\bm x)  |^2$ 
for two values of $\tau_Q$ at various times up to $t = t_{\rm eq}$. These plots help to visualize 
the key point that before $t_{\rm eq}$ when a relative uniform $|\psi|^2$ has not formed one cannot sensibly 
count defects.   \PC{Moreover, it is evident that the defect density is higher for the faster quench.}

\begin{figure}[h!]
\includegraphics[scale = 0.5]{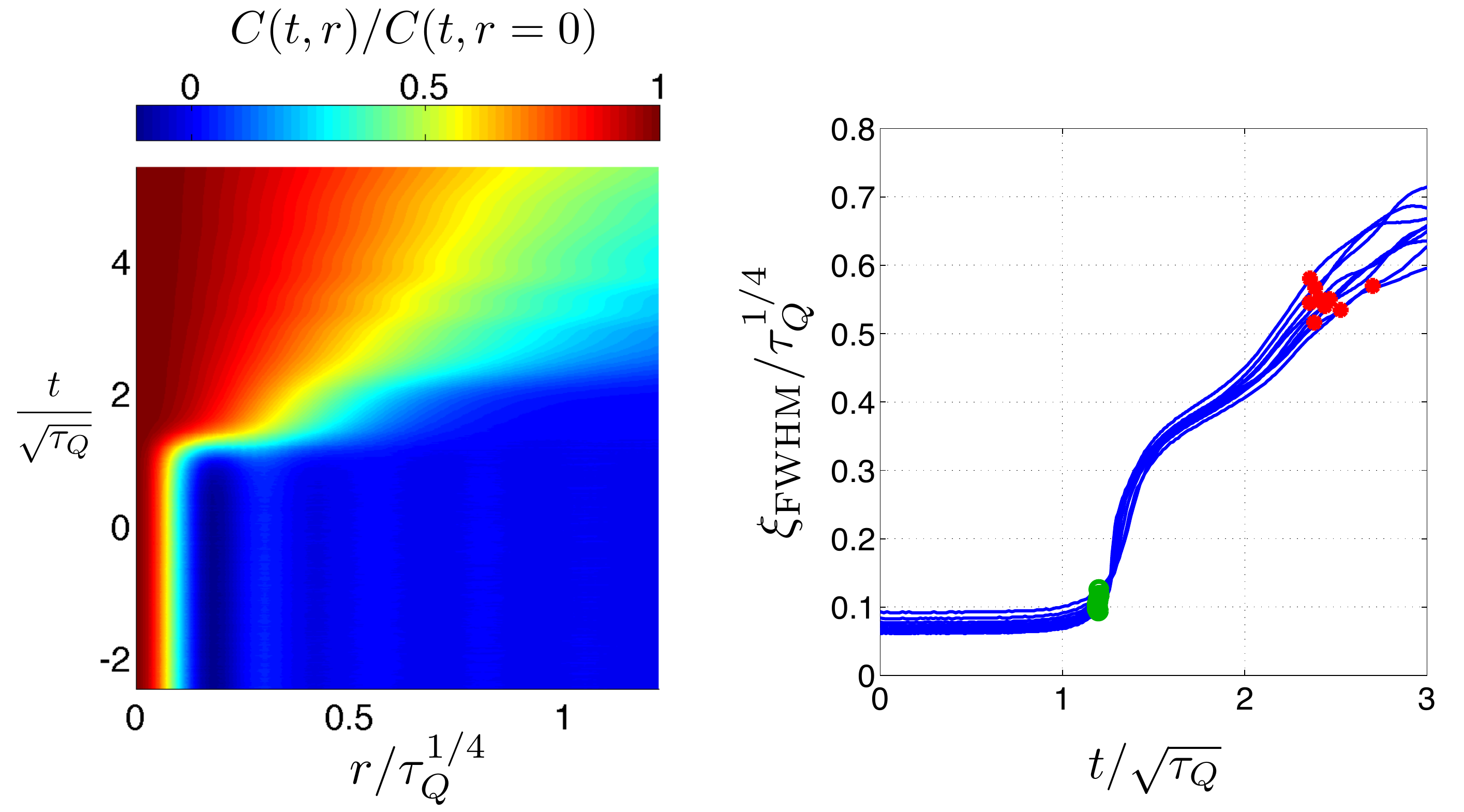}
\caption{{\it Left:}  $C(t,r)$ for $\tau_Q = 4 \tau_o$. 
{\it Right:}  the time evolution of the full width half max $\xi_{\rm FWHM}(t)$ of $C(t,r)$ for $\tau_Q = \tauo n$ with $n = 2,3,4,\dots 10$. The green circles correspond to $\xi_{\rm FWHM}(t_{\rm freeze})$
while the red stars correspond to $\xi_{\rm FWHM}(t_{\rm eq})$.
At $t \approx t_{\rm freeze}$, $\xi_{\rm FWHM}$ starts  a period of growth.  
Note $\xi_{\rm FWHM}(t_{\rm freeze}) \gg \xi_{\rm FWHM}(t_{\rm eq})$.
The collapse of all curves between $t_{\rm freeze}$ and $t_{\rm eq}$ is consistent with the scaling behavior of~\eqref{menx}. It is possible that the observed oscillations superimposed on the square root growth of the $\xi_{\rm FWHM}(t)$ are related to the finite real part of $w_0$ (\ref{bne}).
\label{fig:xi}
} 
\end{figure}

To quantify the time evolution of coarsening and smoothing of the condensate 
we numerically compute the correlation function 
$C(t,r)$ by computing the average in (\ref{eq:Cdef}) over an ensemble of solutions at fixed $\tau_Q$.
The results are in Fig.~\ref{fig:xi}, where 
we also present the full width half max $\xi_{\rm FWHM}(t)$ of $C(t,r)$.  Before $t_{\rm freeze}$, $\xi_{\rm FWHM}$ is dominated by fluctuations and is therefore constant.  After $t \approx t_{\rm freeze}$,
$\xi_{\rm FWHM}$ experiences a period of rapid growth which is consistent with our 
prediction~\eqref{menx} including the scaling behavior. Note
$\xi_{\rm FWHM}(t_{\rm eq})$ is significantly larger than $\xi_{\rm FWHM}(t_{\rm freeze})$, which 
highlights the importance of the ``fudge'' factor needed to account for the correct defect density. 
This is in line with our expectation from Eqs.~(\ref{eq:coarsening}) and (\ref{prd1}).

\begin{figure}[h]
\includegraphics[scale = 0.5]{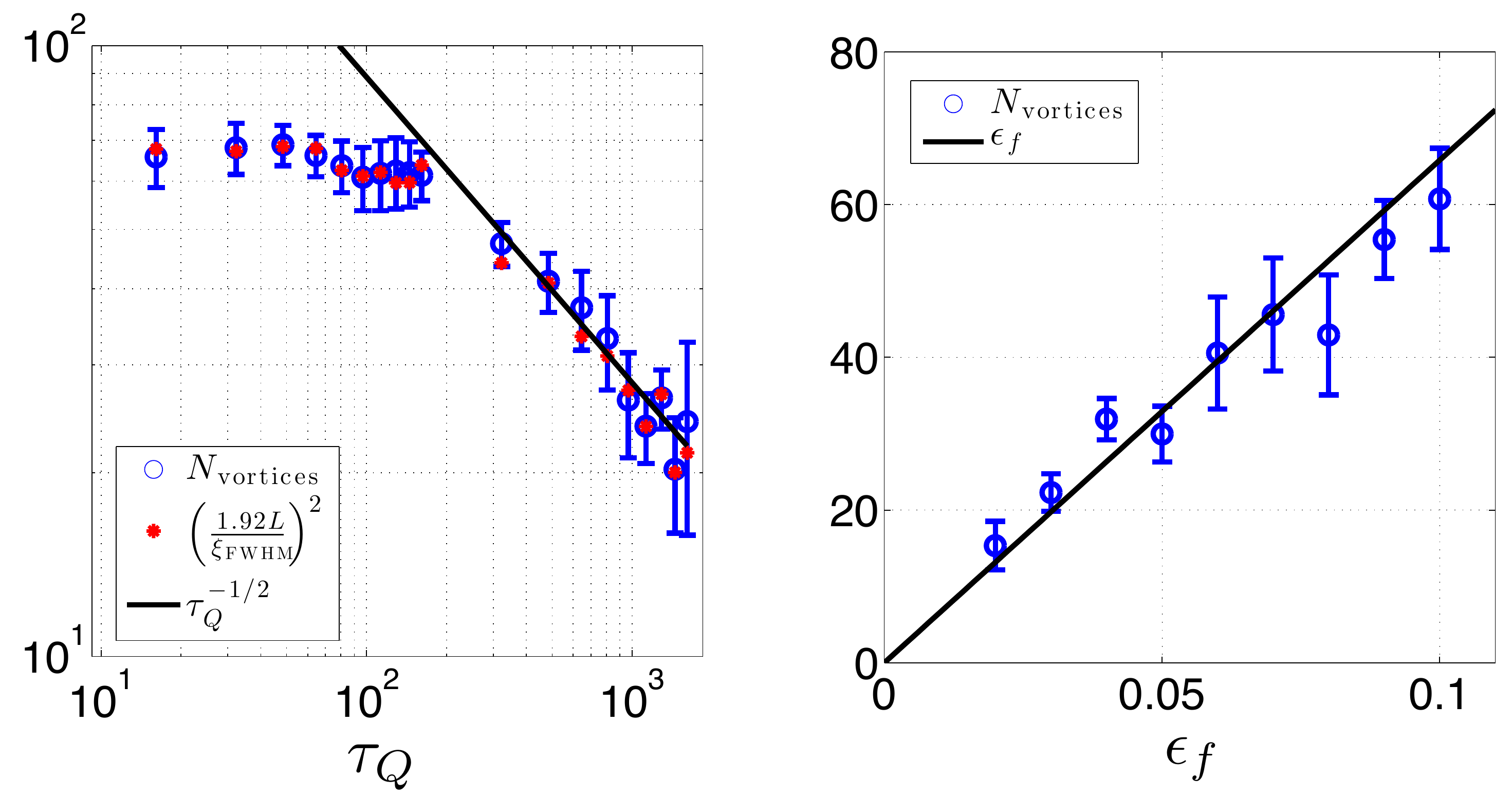}
\caption{{\it Left:} The number of vortices $N_{\rm vortices}$ and $\xi_{\rm FWHM}^{-2}$ at time $t = t_{\rm eq}$ as a function of $\tau_Q$ expressed in units of $T_c$ (\ref{tcmu}). Each data point was computed by averaging the number of vortices over an ensemble of solutions at fixed $\tau_Q$. The error bars were computed from the variance of $N_{\rm vortices}$.   
The numerical results are consistent with KZ scaling $\tau_Q^{-1/2}$ for $\tau_Q > 200$.  For  $\tau_Q < 200$ our numerics are consistent with $N_{\rm vortices} = {\rm const.}$, which is consistent with our expectation that the density of defects should asymptote to a constant in the limit of sudden quenches.
Also included is a plot of $\left (L B/\xi_{\rm FWHM}(t_{\rm eq}) \right )^2$ where $L$ is our box size and $B \approx 1.92$.  
\PC{Our statistics are not sufficient to resolve the logarithmic prefactor in \eqref{prd2}.}
{\it Right:} $N_{\rm vortices}$ versus 
$\epsilon_f = \epsilon(T_f)$, in units of $T_c$ (\ref{tcmu}) for sufficiently small $\tau_Q$.
The results are consistent with~\eqref{rnw} with $N_{\rm vortices} \sim \ep_f$. 
Our statistics are again not sufficient  to resolve the logarithmic prefactor in (\ref{rnw}).
\label{fig:Scaling}
} 
\end{figure}

Finally, in the left panel of Fig.~\ref{fig:Scaling} we show that for slow quenches our numerical results 
reproduce the KZ scaling of the number of vortices $N_{\rm vortices}$.  For  $\tau_Q < 200$ our numerics are consistent with $N_{\rm vortices} = {\rm const.}$
This is the expected behavior from our discussion of the breakdown the KZ scaling in the preceding section: the density of defects should asymptote to a constant in the limit of sudden quenches. For such rapid quenches, the right plot confirms the scaling of the defect density with $\ep_f$ as predicted in~\eqref{den} and~\eqref{rnw}.  
For both situations, our statistics are not enough to resolve the logarithms predicted in~\eqref{prd2} and~\eqref{rnw}. Also included in the left panel of Fig.~\ref{fig:Scaling}
is a plot of $\left (L B/\xi_{\rm FWHM}(t_{\rm eq}) \right )^2$ where $B \approx 1.92$. The  fantastic agreement between 
$\left (L B/\xi_{\rm FWHM}(t_{\rm eq}) \right )^2$ and $N_{\rm vortices}$ for all 
$\tau_Q$ bolsters the notion that the vortex density is a measure of the correlation length.
Moreover, the observation that $B = O(1)$ and $\xi_{\rm FWHM}(t_{\rm eq}) \gg \xi_{\rm FWHM}(t_{\rm freeze})$ is consistent with our argument that coarsening during the early stages of the evolution can dramatically increases the correlation length and decreases the expected density of defects from the KZ prediction (\ref{eq:KZdefectdensity}).

\section{Conclusion and discussion}

To summarize, we elucidated a novel period of non-adiabatic evolution 
after a system passes through a second order phase transition, where a 
parametrically large amount of coarsening occurs \textit{before} a well-defined  
condensate forms.  \PC{The physical origin of the coarsening can be traced to the fact that 
when the system passes through the phase transition, IR modes of the order parameter become unstable and exponentially grow.}
We showed that such a far-from-equilibrium coarsening regime
could have important consequences for defect formation. We also numerically simulated 
thermal quenches in a $2+1$ dimensional holographic superfluid, which provided strong support 
for our analytic results. 

For slow quenches a key quantity we introduced is $R$ of~\eqref{defR} which we copy here for convenience
\be 
R  \sim \fluc^{-1} \tau_Q^{{\Lam \ov 1+ \nu z}}, \qquad \Lam \equiv {(d - z)\nu -2 \beta}  \ .
\ee 
For $R \gg 1$, there is large hierarchy between $t_{\rm freeze}$ and $t_{\rm eq}$, and the density of 
defects can be significantly lower than that predicted by KZ. Systems with $R \gg 1$ can be 
separated into two cases:

\ben 

\item[I.]  The exponent $\Lam$ is positive, i.e. 
\be \label{oem}
(d - z)\nu > 2 \beta 
\ee
for which $R \to \infty$ as  $\tau_Q \to \infty$. For mean field with $z=2, \nu = \ha, \beta = \ha$, this implies $d > 4$, i.e. above the upper critical dimension of mean field theory. 
Using hyperscaling relation $ 2 \beta =  (d-2 + \eta) \nu$, for a general critical point $\Lam$ can be simplified as 
\be 
\Lam = (2-\eta-z) \nu 
\ee
and~\eqref{oem} becomes
\be 
 z < 2 - \eta \ . 
\ee
An example which satisfies this condition is superfluid $^4{\rm He}$ which has 
\be 
z = {3 \ov 2}, \qquad \eta \approx 0.037 \ .
\ee
Other examples include three-dimensional isotropic antiferromagnet and the three-dimensional XY model.


\item[II.] $\Lam$ is negative, but $\fluc \ll 1$ so that for large but finite $\tau_Q$ we still have $R \gg 1$, i.e. 
\be \label{conde}
\fluc \ll \tau^{{\Lam \ov 1+ \nu z}}_Q  \ .
\ee
\PC{One class of examples is holographic theories, such as that  discussed in Sec.~\ref{sec:gravdes}, for which $\fluc \sim {1 \ov N^2}$ with $N \to \infty$.}
As another class of examples, let us consider model A for dynamic critical phenomena~\cite{Hohenberg}.
Recall that the Van Hove theory of critical slowing down predicted exactly  $z=2 -\eta$. Renormalization group analysis give $z$ slightly greater than this value~\cite{folk2006critical}, which means 
that generically for model A, $\Lam$ is only slightly negative, and thus~\eqref{conde} essentially translates 
into $\fluc \ll 1$. As an explicit example, conventional superconductors have a very small $\fluc$ and thus we expect them to have a large hierarchy between $t_{\rm freeze}$ and $t_{\rm eq}$.

\een

For fast quenches the analogous quantity is $R_f$ defined in~\eqref{bmme}.  
Comparing~\eqref{bmme} with~\eqref{defR} we see the conditions for $R_f \gg 1$ are essentially 
identical to those for $R \gg 1$, and the above discussion also applies.

\acknowledgments

We thank Laurence Yaffe for useful discussions.
The work of PC 
is supported by the Fundamental Laws Initiative of the
Center for the Fundamental Laws of Nature at Harvard
University.  The work of HL is partially supported by the U.S. Department of Energy (D.O.E.) under cooperative research agreement \#DE-FG0205ER41360. 
AMG was supported by EPSRC, grant No. EP/I004637/1, FCT, grant PTDC/FIS/111348/2009
and a Marie Curie International Reintegration Grant PIRG07-GA-2010-268172.

\appendix

\section{Details on various integrals} \label{app:b}

In this section we give some details in the derivation of~\eqref{eq:Cfourierspace} and~\eqref{apme}.  

\subsection{Slow quenches} 

Consider the integral~\eqref{eq:Cint} for small $q \lesssim q_{\rm max}$ 
\bea
\label{eq:Cint1}
C(t,q) &=&  \int_{t_{\rm freeze}}^t  dt' \fluc |H(q)|^2  e^{2  \int^t_{t'} dt'' {\rm Im} \, \w_0(\epsilon(t''),q)}   \cr
& \approx &  \fluc  |H(0)|^2 t_{\rm freeze} \int_{1}^{\bar t}  d \bar t'   \exp \le[ 
2   t_{\rm freeze} \int^{\bar t}_{\bar t'} d \bar t'' \, {\rm Im} \, \w_0(\epsilon (t''),q)  \ri]
\eea
where we have introduced $\bar t = t/t_{\rm freeze}$. Now note from~\eqref{eq:quasinormaldisp} that 
\be 
\label{eq:sca1}
t_{\rm freeze} {\rm Im} \, \w_0 (\epsilon (t),q) = - {a \tau_s \ov \xi_s^2} \bar t^{(z - 2) \nu} q^2 \xi_{\rm freeze}^2 +  b \tau_s \bar t^{z \nu} + \dots,
\ee
where we have used~\eqref{eq:linquench}--\eqref{eq:xiandtfreeze}. We thus find that 
\be 
C(t,q) \approx \fluc  C |H(0)|^2 t_{\rm freeze} \exp \le(a_2 \bar t^{1 + \nu z} - \ha q^2 \ell_{co}^2 (\bar t) \ri)  , \qquad
\ell_{co}^2 (\bar t) \equiv a_3^2 \, \xi_{\rm freeze}^2 \bar t^{1 + (z-2) \nu}
\ee
where 
\be 
a_2 = {2b \tau_s \ov 1 + z \nu}, 
\quad a_3^2 = 
{4a \tau_s \ov (1 + \nu (z-2)) \xi_s^2},
\quad C = 
\int_{1}^{\bar t}  d x \, \exp \le(- a_1 x^{1+\nu z} + \ha a_3^2 q^2 \xi_{\rm freeze}^2 x^{1 + \nu (z - 2) } \ri)  \ .
\ee
Note that since $x > 1$, $\nu>0$  and we are interested in the regime $q \xi_{\rm freeze} \lesssim O(1)$,  
the first term in the exponential of $C$ always dominate over the second term. For large $\bar t$, due to exponential suppression, the integral for $C$ is dominated by the lower end, and we thus conclude that 
$C \sim O(1)$. For $\bar t \sim O(1)$, the $\bar t$ dependence is more complicated, but this is not the regime we are interested in. Suppressing various $O(1)$ prefactors we thus find~\eqref{eq:Cfourierspace}--\eqref{eq:coarsening}.  

Note that in~\eqref{eq:Cint} we have assumed the condensate starts growing at $t_{\rm freeze}$, but it is clear from our derivation that~\eqref{eq:Cfourierspace}--\eqref{eq:coarsening} are not sensitive to the specific
time when the condensate starts growing. For example, the conclusion remains the same if the lower end 
of the integral of~\eqref{eq:Cint1} is changed to $0$.

\subsection{Rapid quenches} 

For rapid quenches discussed in Sec.~\ref{sec:iic} we have 
\be
\label{eq:Cint3}
C(t,q) =  \int_{t_{\rm freeze}}^t  dt' \fluc |H(q)|^2  e^{2 {\rm Im} \, \w_0(\epsilon_f ,q) (t - t')}  
\approx  {\fluc |H(0)|^2 \ov  {\rm Im} \, \w_0(\epsilon_f ,q)} \le[e^{2 {\rm Im} \, \w_0(\epsilon_f ,q) (t - t_{\rm freeze})} - 1 \ri] \ .
\ 
\ee
Given the scaling form of ${\rm Im} \, \w_0 (T, q) = \ep^{\nu z} {\rm Im} \, h(q \ep^{-\nu})$, the above equation can be written as 
\be 
C(t, q) \approx  {\ep_f^{-\nu z} \fluc |H(0)|^2 \ov 2 {\rm Im}\,h(\tilde q)} \le[e^{2 {\rm Im}\, h (\tilde q) \tilde t} -1 \ri] , \quad
\tilde q \equiv q \ep_f^{-\nu}, \quad \tilde t \equiv (t - t_{\rm freeze}) \ep_f^{\nu z}
\,
\ee
whose Fourier transform can also be written in a scaling form 
\be\label{nmm3}
C(t,r) = \ep_f^{(d-z) \nu} \fluc  f(\tilde t,\tilde r ), \qquad \tilde r \equiv  r  \ep_f^\nu
\ee
for some scaling function $f$. For large $\tilde r \gg 1$, we can use the small $\tilde q$ expansion 
$h (\tilde q) = b - a \tilde q^2$, and for $\tilde t \gg 1$, we find that  
\be \label{apmeq}
C (t, r) \sim \ep_f^{(d-z) \nu} \fluc  \exp \le[2 b  (t - t_{\rm freeze}) \ep_f^{\nu z} - {r^2 \ov 2 \ell_{\rm co}^2 (t)} \ri] 
\ee
with 
\be 
\ell_{\rm co}^2 (t) = 4a  (t - t_{\rm freeze}) \ep_f^{\nu (z-2)} \ . 
\ee

%

\section{Non-equilibrium holographic superfluidity: gravity setup}\label{app:a}

The field content of the $2+1$ dimensional holographic superfluid we study consists of the metric $G_{MN}$, a $U(1)$ gauge field $A^M$ and a charged scalar $\Phi$ with charge $e$.  These fields live in asymptotically AdS$_4$ spacetime. 
Following \cite{Hartnoll:2008vx} we take the action to be
\begin{equation}
\label{eq:action}
S_{\rm grav} = \frac{1}{16 \pi G_{\rm Newton}} \int d^4 x \sqrt{-G} \left [ R + \Lambda + \frac{1}{e^2} \left ( - \frac{1}{4} F_{MN}F^{MN} - | D \Phi|^2 - m^2 |\Phi|^2 \right ) \right ],
\end{equation}
where $R$ is the Ricci scalar, $F_{MN}$ is the $U(1)$ field strength and $D$ is the gauge covariant derivative and $G = - \det G_{MN}$.   The mass $m$ of the scalar field and the cosmological 
constant $\Lambda$ are given by
\begin{align}
m^2 &= -2  &\Lambda &= -3.
\end{align}
The $U(1)$ gauge redundancy in the bulk
encodes a $U(1)$ global symmetry in the boundary theory where  
the boundary order parameter transforms with a phase $ \psi  \to  \psi  e^{i \alpha}$.  Indeed, the 
bulk scalar field $\Phi$ encodes $ \psi $.

Following \cite{Chesler:2013lia} we employ infalling Eddington-Finkelstein coordinates where the metric takes the form
\begin{equation}
ds^2 = r^2 g_{\mu \nu}(t,\bm x,r) dx^\mu dx^\nu + 2 dr dt.
\end{equation}
Here Greek indices run over boundary spacetime coordinates and $r$ is the AdS
radial coordinate with $r = \infty$ the AdS boundary.  With our choice of coordinates lines of constant $t$
represent infalling null radial geodesics affinely parameterized by $r$.  
In addition we choose to work in the gauge $A_r = 0$.

For simplicity we choose to work in the probe limit $e \to \infty$ where gravitational
dynamics decouple from the dynamics of the gauge and scalar fields.  The equations of motion following from (\ref{eq:action}) are then simply
\begin{subequations}
\label{eq:eom}
\begin{eqnarray}
\label{eq:eomEin}
0 &=& R_{MN} - \frac{1}{2} G_{MN} (R + 2 \Lambda ) ,
\\ \label{eq:eomMax}
0 & = &\nabla_M F^{NM} - J^M, 
\\ \label{eq:eomKG}
0 &=& (-D^2 + m^2) \Phi.
\end{eqnarray}
\end{subequations}

Since the boundary of AdS is time-like, the equations of motion (\ref{eq:eom})
require boundary conditions to be imposed there.
As the boundary geometry of AdS corresponds to the geometry the dual quantum theory lives in,
we demand that the boundary geometry be that of flat $2+1$ dimensional Minkowski space.  
This is accomplished by setting $\lim_{r \to \infty} g_{\mu \nu} = \eta_{\mu \nu}$.
The near-boundary behavior of the gauge and scalar fields can easily be worked by 
from Eqs.~(\ref{eq:eomMax}) and (\ref{eq:eomKG}) and read
\begin{eqnarray}
A_{\alpha}(t,{\bm x},r) &=& A_{\alpha}^{(0)}(t,\bm x) + \frac{A_{\alpha}^{(1)}(t,\bm x)}{r} + O(1/r^2),
\\ \label{eq:scalarasm}
\Phi(t,{\bm x},r) &=& \frac{\Phi_{(1)}(t,{\bm x}) }{r}  + \frac{\Phi_{(2)}(t,{\bm x})}{r^2}  + O(1/r^3).
\end{eqnarray}
On the gauge field we impose the boundary condition
\begin{equation}
A_{\alpha}^{(0)}(t,\bm x)  = \delta_{\alpha 0} \mu,
\end{equation}
where $\mu$ is a constant.  In the dual QFT $\mu$ is interpreted as a chemical potential for the conserved $U(1)$ charge. 
As a final boundary condition we set 
\begin{align}
\label{eq:Phibc}
\Phi_{(1)}(t,{\bm x}) &= \varphi(t,{\bm x}).
\end{align}
with $\varphi$ random variable satisfying statistics (\ref{nne}).
The stochastic driving of the scalar field mimics the effect of quantum and thermal fluctuations
induced by the black brane's Hawking radiation.  In the dual quantum theory the boundary condition 
(\ref{eq:Phibc}) amounts to deforming the Hamiltonian 
\begin{equation}
H \to H + \int d^2x \left \{\varphi^* \psi + \varphi \psi^* \right \}.
\end{equation}
Note $\varphi$ has mass dimension one and $\psi$ has mass dimension two.
In terms of the asymptotic behavior of the scalar field (\ref{eq:scalarasm})
the boundary order parameter reads
\begin{equation}
\label{eq:bulkandboundary}
 \psi(t,\bm x)  = \Phi_{(2)}(t,{\bm x}) - (\partial_t- i \mu) \varphi(t,{\bm x}).
\end{equation}

Let us first discuss static equilibrium solutions to the set of equations of motion (\ref{eq:eom}).
Translationally invariant equilibrium solutions to Einstein's equations consist of 
black branes,
\begin{equation}
\label{eq:geoeq}
ds^2 =  r^2 \left [ -f dt^2 + d {\bm x}^2 \right] + 2 dr dt,
\end{equation}
where 
\begin{equation}
f =1 - \left (\frac{r_h}{r} \right)^3.
\end{equation}
The Hawking temperature $T$ of the 
black brane is related to the horizon radius $r_h$ by 
\begin{equation}
 r_h = \frac{4 \pi T}{3},
\end{equation}
and corresponds to the temperature of the dual quantum theory.

Static equilibrium solutions to the scalar-gauge field system (\ref{eq:eomKG}) and (\ref{eq:eomMax}) 
were first explored in \cite{Hartnoll:2008vx}.
One static solution to (\ref{eq:eomKG}) and (\ref{eq:eomMax}) 
(with $\varepsilon = 0$ and hence no stochastic driving) is simply 
\begin{subequations}
\label{eq:eqsols}
\begin{eqnarray}
\label{eq:Aeqsol}
A_\alpha &=& \mu \left ( 1 - \frac{r_h}{r} \right ) \delta_{\alpha 0},
\\
\Phi &=& 0.
\end{eqnarray}
\end{subequations}
However, for sufficiently low 
temperatures this solution is unstable and not thermodynamically preferred.  For 
$T < T_c$, where 
\begin{equation}
T_c \approx 0.0587 \mu,\label{tcmu}
\end{equation}
the thermodynamically preferred solution has $\Phi \neq 0$.  Hence the bulk $U(1)$ gauge redundancy is 
spontaneously broken at low temperatures and the black brane develops a charged scalar atmosphere.  
Likewise, via (\ref{eq:bulkandboundary}) the boundary order parameter is non-zero
and the global $U(1)$ symmetry on the boundary is spontaneously broken.  The gravitational and boundary systems
have a second order phase transition at $T = T_c$ with mean-field critical exponents.

To study the Kibble-Zurek mechanism gravitationally we drive the system stochastically with the boundary condition
(\ref{eq:Phibc}) and choose to dynamically cool the black brane geometry
through $T_c$.  When the geometry cools through $T_c$ the aforementioned instability 
will result in the scalar field $\Phi$ growing and the black brane developing a scalar atmosphere.
Likewise, as this happens the boundary QFT condensate (\ref{eq:bulkandboundary}) will grow in amplitude.

Instead of solving Einstein's equations 
(\ref{eq:eomEin}) for a black brane with dynamic temperature, we chose to fix the geometry to be the equilibrium geometry (\ref{eq:geoeq})
but with a time dependent temperature $T(t)$ equal to the boundary quench protocol temperature (\ref{eq:linquench}), which we control.
The metric will therefore no longer satisfy Einstein's equations.
Why is it reasonable to employ a geometry that does not satisfy Einstein's equations?
To answer this question we note that to cool the system through $T_c$ we can 
couple it to an external thermal bath at controllable temperature $T_{\rm ext}(t)$.
This can be done by, for example, putting our system in a box of size $L$ and putting the surface of the 
box in contact with the thermal reservoir.   As we are ultimately interested in slow quenches 
where $T_{\rm ext}'(t)$ is parametrically small, we expect thermal equilibration 
and $T(t) \approx T_{\rm ext}(t)$.  In this limit Einstein's equations can be solved with 
the gradient expansion of fluid/gravity \cite{Bhattacharyya:2008jc}.
At leading order in gradients the solution is simply (\ref{eq:geoeq}), but with the time dependent temperature $T(t)$.

Our numerical methods used to solve the scalar/gauge field system (\ref{eq:eomKG}) and (\ref{eq:eomMax})
are outlined in \cite{Chesler:2013lia}.  We use pseudospectral methods and discretize the AdS radial coordinate using 20  
Chebyshev polynomials.  In the spatial directions we work in a periodic spatial box and discretize using a basis of 201 plane waves.
We chose box size $L T_c = 30.8$ and measure all other dimensionful quantities in units of $T_c$.  We choose noise amplitude $\fluc T_c = 1.5 \times 10^{-3}$.
As our quench protocol (\ref{eq:linquench}) 
starts off at temperatures $T> T_c$, in the infinite past we choose initial conditions (\ref{eq:eqsols}).

\bibliography{refs}%
\end{document}